\documentclass[12pt]{article}

\usepackage{graphicx}
\usepackage{dcolumn}
\usepackage{bm}
\usepackage{upgreek}
\usepackage{amsmath} 

\begin{document}

\vspace{1.5cm}

\begin{center}
{\Large \bf  Improved cosmological bounds for a fine-tuned see-saw mechanism of keV sterile neutrinos \hfill\\}
\end{center}
\vspace{0.5cm}

\begin{center}
{M.N. Dubinin$^{a}$, D.M. Kazarkin$^{b}$\footnote{corresponding author, e-mail: {\tt kazarkin.dm17@physics.msu.ru}} \\
\hfill\\
{\small \it $^a$Skobeltsyn Institute of Nuclear Physics, Lomonosov Moscow State University}\\
{\small \it 119991, Moscow, Russia} \\
{\small \it $^b$Physics Department, Lomonosov Moscow State University}\\
{\small \it 119991, Moscow, Russia}
}
\end{center}

\vspace{1.0cm}
\begin{center}
{\bf Abstract}
\end{center}
\begin{quote}

Considering the gauge model $SU(2)_L \times U(1)$ with the extension of the lepton sector by right-handed sterile Majorana neutrinos, we distinguish a "minimal parametric mixing scenario" for a complete matrix of $6 \times 6$ light (active) and sterile neutrinos, which depends, in addition to experimentally measured physical parameters, only on the masses of sterile neutrinos. For such a scenario, improved cosmological bounds are provided, resulting from the lifetime of sterile neutrinos and the fraction of energy carried by sterile neutrino dark matter.
\end{quote}   

\section{\label{sec:introduction}Introduction}

The Standard Model (SM) of particle physics, confirmed with high accuracy in experiments, does not explain a number of observed phenomena, one of which is neutrino oscillations detected in numerous experiments \cite{general1} and testifying to the existence of neutrino masses and mixing between three generations of neutrinos. The available data are consistent with the oscillations between $\nu_e$, $\nu_{\mu}$ and $\nu_{\tau}$, which mass splitting can be implemented within the framework of two mass hierarchies. In the case of a normal hierarchy, $m_1\ll m_2<m_3$, $m_2\approx\sqrt{\Delta m^2_{sol}} \sim 8,6\times 10^{-3}$ eV, $m_3\approx\sqrt{\Delta m^2_{atm}+\Delta m^2_{sol}} \sim 0.05$ eV where $\Delta m^2_{atm} \equiv |\Delta m^2_{32}| = 2.5 \times 10^{-3}$ eV and  $\Delta m^2_{sol} = \Delta m^2_{21} =7.4 \times 10^{-5}$ eV \cite{pdg}.
Extension of the SM leptonic sector by three right-handed Majorana fermions ${\nu_a}_R$, singlets relative to the gauge group $SU(2)_L\times U(1)$ \cite{sterile1,sterile2}, is a natural and rather simple way to include neutrino masses. The corresponding Lagrangian of the extension has the form
\begin{equation}\label{lagr}
	\mathcal{L}_\nu = i \overline{\nu_{a}}_R \gamma^\mu\partial_\mu {\nu_{a}}_R - \left(F_{la}\overline{L_l}{\nu_{a}}_R\tilde{\Phi} + \frac{1}{2}{(M_M)}_{ab}~\overline{\nu^C_{a}}_R \, {\nu_{b}}_R + h.c.\right),	
\end{equation}
where $L_l^T=(\nu_l,l)_L$, $l=e,\mu,\tau$ is the left SM doublet, $\tilde{\Phi}$ is the conjugate Higgs doublet, $a,b=1,2,3$, $F_{la}$ is 3$\times$3 matrix of Yukawa couplings, $M_M$ is Majorana mass matrix. $SU(2)$-states $\nu_{l_L}$ are phenomenologically related to mass states $\upnu_i$ by the Pontecorvo-Maki-Nakagawa-Sakata (PMNS) mixing matrix \cite{pmns}
\begin{equation}
\label{pmns}
    \nu_{l_L}=  U_{li}\upnu_{i_L}.
\end{equation}
The smallness of the masses of SM neutrinos (or active neutrinos) compared to charged leptons of the corresponding generation is naturally explained by the see-saw mechanism \cite{seesaw1, seesaw2, seesaw3}. In the simplest case of see-saw type 1, the mass matrix of active neutrinos has the form
\begin{eqnarray}
m_{\nu} = - M_{D} M^{-1}_{M} M_{D}^{T},
\label{eq:seesaw}
\end{eqnarray}
where $(M_{D})_{ia}=F_{ia}\, v$ is Yukawa matrix after spontaneous symmetry breaking (or matrix of Dirac mass terms), 
$v=\sqrt{2} <\Phi>$. 
Difference of the eigenvalues for $M_{M}\gg M_{D}$ provides small masses of active neutrinos, whereas the masses of HNLs can vary greatly. For example, for the Yukawa constants $F\sim 10^{-7} - 1$ for the mass HNL $M_M\sim 10^{2} - 10^{15}$ GeV $m_{\nu}\sim 10^{-2}$ eV. Methods of experimental detection of HNL vary depending on the scale of their masses, for example, GeV or keV. Difficulties in isolating signals at colliders can be overcome in experiments such as SHiP \cite{alekhin}.

The case with the mass parameter $M_M$ of the order of keV is of interest, since the lightest HNL may be a good candidate for the role of dark matter \cite{hnl_dm}. Direct astrophysical constraints connected with $\gamma$ radiation in HNL decays are consistent with $M_M\sim$10 keV \cite{boyarski,Merle:2013gea}.
The minimal extension of the lepton sector of the SM, the $\nu$MSM model \cite{shaposhnikov1,shaposhnikov2}, includes three generations of sterile neutrinos, one of which is a candidate for the role of dark matter particle at keV mass scale, while heavier HNLs provide small masses of active neutrinos due to see-saw mechanism.
Significant limitations of the model parameter space can be imposed by cosmological observations, which lead to at least three HNL and establish a strict upper limit on the mass of the lightest active neutrino.
Experiments on neutrino oscillations give differences in the squares of the masses of active neutrinos, not absolute values of their masses, therefore, for three generations of leptons there are 18 very weakly limited model parameters (3 parameters of the Dirac mass term, 3 parameters of the Majorana mass, 6 mixing angles and 6 phases) for both mass hierarchies in the case of independent mixing in the sectors of active neutrinos and HNL, 9 of which (PMNS parameters and three masses of active neutrinos) can be determined experimentally at present time.
The masses of active neutrinos are limited by observations of the large-scale structure \cite{activenu}, which give an upper limit for the sum $\sum m_i$ and set an upper bound $\Omega_\nu h^2$ for the energy fraction of active neutrinos in the Universe $\Omega_{DM}h^2 =$0.1186$\pm$0.0020 \cite{pdg}.

Currently, to describe the searches of HNLs the so-called ''model-independent phenomenological approach'' is used assuming that only a single HNL is available in experiments with extracted beams or in colliders, while other HNLs are sufficiently heavy and decouple not affecting the analysis. There are only two {independent} parameters in this approach, the HNL mass and the Yukawa coupling defining HNL interaction with an active neutrino of a given flavor, assuming that the mixing with other flavors is zero. Such simplification is useful for derivation of generic bounds on the mixing parameter beyond any aspects of a particular model construction, but needs an appropriate translation if one would like to go beyond the case of one generation and consider a well-defined mixing.
In the following, diagonalization \cite{ibarra1,ibarra2} of the lepton sector of the $\nu$MSM model is used for the realistic case of three HNL generations. In this sense, the model under consideration is a generalization of $\nu$MSM to the case of an arbitrary non-diagonal mixing matrix in the sector of heavy neutral leptons.

\section{\label{sec:section2}Mass states and neutrino mixing}
The full mass matrix $\mathcal{M}$ $6\times6$ described by the Lagrangian \eqref{lagr} is a complex symmetric matrix that be represented as $\mathcal{M}=\mathcal{U}\mathcal{D}\mathcal{U}^T$ (so-called Takagi factorization), where  $\mathcal{U}$ is a unitary matrix, $\mathcal{D}$ is a diagonal non-negative matrix
    \begin{eqnarray*}
        \mathcal{U}^\dagger
        \left( \begin{array}{cc}
        0 & M_{D} \\
        M_{D}^{T} & M_{M} 
        \end{array} \right)
         \mathcal{U}^* = 
        \left( \begin{array}{cc}
        \hat{m} & 0 \\
        0 & \hat{M} 
        \end{array} \right)
    \end{eqnarray*}
\\
where $\hat{m}=diag(m_{1},m_{2},m_{3})$ is the diagonal matrix of active neutrino masses, and $\hat{M}=diag(M_{1},M_{2},M_{3})$ -- diagonal mass matrix for HNL states $N_{i}$. Taking  $\mathcal{U}$ in the form
    \begin{equation}
        \mathcal{U}= \mathcal{W} \left(\begin{array}{cc}
        U_\nu & 0 \\
        0 & U_N^* 
        \end{array}\right)
    \end{equation}
it follows from the unitarity condition for $\mathcal{U}$ that $\mathcal{W}, U_\nu$ and $U_N$ are unitary matrices.
In the following the diagonalization from \cite{ibarra1} is used
    \begin{eqnarray*}
        \mathcal{W} = exp
        \left( \begin{array}{cc}
        0 & R \\
        -R^{\dagger} & 0
        \end{array} \right)
        =\left( \begin{array}{cc}
        1-\frac{1}{2}RR^{\dagger} & R \\
        -R^{\dagger} & 1-\frac{1}{2}R^{\dagger}R
        \end{array} \right) + \mathcal{O}(R^{3},R^{4}),
    \end{eqnarray*}
where $R$ is $3\times 3$ complex matrix. It is assumed that the eigenvalues of $R$ are small. The matrix $U_\nu$ is defined in the sector of active neutrinos so that $U_\nu^{T}m_{\nu}U_\nu =\hat{m}$, and in the sector of charged leptons a basis with canonical ordering is taken where analogous matrix $U_{l}$ is equal to unity matrix\footnote{In the case of non-canonical ordering of charged leptons, this matrix should be taken as a permutation matrix $U_{l}=P_{3\times3}$, as, for example, in \cite{discretesym1}}.
The Lagrangian terms for active neutrino currents have the form
    \begin{eqnarray} \label{lagr:act currents}
        \mathcal{L}^\upnu_{CC} &=& -\frac{g}{\sqrt{2}}\bar{l_L}\gamma_{\mu}U_{\rm PMNS}\upnu_{i_L}W^{\mu}+h.c.
        \label{lagr:act currents:cc} 
        \\
        \mathcal{L}^\upnu_{NC} &=& \frac{g}{2c_{W}}\bar{\upnu}_{i_L}U_\nu^\dagger\gamma_{\mu}(I+\eta+\eta^\dagger)U_\nu\upnu_{j_L}Z^{\mu}+h.c.
        \label{lagr:act currents:nc}
    \end{eqnarray}
where $\upnu_{iL}$ are the mass states of active neutrinos, $W$ and $Z$ are the fields of SM vector bosons. The parameter $\eta=-\frac{1}{2}RR^\dagger$ characterizes the deviation of the PMNS matrix from untarity, $(I+\eta)U_\nu=U_{\rm PMNS}$.
Charged and neutral HNL currents have the form
    \begin{eqnarray} \label{lagr:hnl currents}
       \mathcal{L}_{CC}^N &=& -\frac{g}{\sqrt{2}}\bar{l_L}\gamma_{\mu}(R U_N^*)_{lk}N_{k_L}W^{\mu}+h.c.
       \label{lagr:hnl currents:cc}
       \\
       \mathcal{L}_{NC}^N &=& \left(-\frac{g}{2c_{W}}\bar{\upnu}_{i_L}\,  \gamma_{\mu}U_\nu^\dagger(I+\eta)(R U_N^*)N_{j_L}Z^{\mu} + h.c.\right) - \\ \nonumber
       &&-\frac{g}{2c_{W}}\overline{N}_{i_L}\gamma_\mu U_N^T R^\dagger R U_N^*N_{j_L}Z^{\mu}.
       \label{lagr:hnl currents:nc}
    \end{eqnarray}
Thus, the mixing of HNL and the left active neutrino is characterized by the matrix $\Theta \equiv R U^*_N$. Further on it is assumed that $N_{i}$ have sufficiently large masses, so that the eigenvalues of $M_{N}$ are greater than the eigenvalues of $M_{D}$. Using the $\mathcal{W}$-matrix decomposition, one can explicitly write a set of equations for diagonalization of the mass matrix
    \begin{eqnarray} 
        \left( \begin{array}{cc}
        1-\frac{1}{2}RR^{\dagger} & -R \\
        R^{\dagger} & 1-\frac{1}{2}R^{\dagger}R
        \end{array} \right) \nonumber
        \left( \begin{array}{cc}
        0 & M_{D} \\
        M_{D}^{T} & M_{M} 
        \end{array} \right)\left( \begin{array}{cc}
        1-\frac{1}{2}R^{*}R^{T} & R^{*} \\
        -R^{T} & 1-\frac{1}{2}R^{T}R^{*}
        \end{array} \right) \\=\left( \begin{array}{cc}
        U_\nu\hat{m}U_\nu^{T} & 0 \\
        0^{T} & U_N^{*}\hat{M}U_N^{\dagger} 
        \end{array} \right) 
    \end{eqnarray}
Keeping the terms up to $\mathcal{O}(R^2)$ 
    \begin{eqnarray}
        M_{D}-R M_{M} + O(R^2) &\simeq& 0,    \label{eq:R1} 
        \\
        -M_{D}R^{T}-R M_{D}^{T}+R M_{M}R^{T} + O(R^3) &\simeq& U_\nu \hat{m}U_\nu^{T} \equiv m_{\nu},  \label{eq:R2} 
        \\
        M_{M}+R^{\dagger}M_{D}+M_{D}^{T}R^* + O(R^2) &\simeq& U_N^{*}\hat{M}U_N^{\dagger} \equiv M_N, \label{eq:R3}
    \end{eqnarray}
from where in the approximation $RR^\dagger \approx 0$
    \begin{eqnarray}
        R &=& M_{D}M^{-1}_{M}, 
        \\
        m_{\nu} &=& -RM_{M}R^T,
    \end{eqnarray}
and this in turn leads to the expression \eqref{eq:seesaw}. The equations \eqref{eq:R2} and 
\eqref{eq:R3} can be used to obtain constraints on the elements of the matrix $\Theta=R U_N^*$ using experimental bounds on the effective masses of active neutrinos
    \begin{eqnarray}
        \sum_{k} |\Theta_{l^\prime k} {\hat M}_{k} \Theta^{T}_{kl}| \le (m_\nu)_{l^\prime l}, \quad l^\prime,l = e,\mu,\tau.
        \label{eq:restriction}
    \end{eqnarray}
Rewriting \eqref{eq:seesaw} taking into account \eqref{eq:R3}  
    \begin{eqnarray}
        \label{eq:R6}
        U_N^* \hat{M} U_N^\dagger = M_{N} \simeq M_M = -M_{D}^{T} m_{\nu}^{-1} M_{D} = - M_{D}^{T} U_\nu^{*} \hat{m}^{-1} U_\nu^{\dagger} M_{D}
    \end{eqnarray}
we get the equivalent form
    \begin{equation}
        I = \Omega^{T}\Omega = [-i\sqrt{\hat{m}^{-1}} U_\nu^{\dagger} M_{D} U_N \sqrt{\hat{M}^{-1}} ]^{T}  [-i\sqrt{\hat{m}^{-1}} U_\nu^{\dagger} M_{D} U_N \sqrt{\hat{M}^{-1}}]
        \label{eq:orth_condition}
    \end{equation}
where $\Omega$ denotes an arbitrary orthogonal matrix. 
The matrix $\Omega$ can be parameterized by three (generally speaking, complex) parameters. For example, using the three Euler angles $\alpha_1, \alpha_2, \alpha_3$
    \begin{equation}
        \label{Omega:Euler}
    	\Omega({\alpha_1}, {\alpha_2}, {\alpha_3})=\left(
    	\begin{array}{ccc}
    		c _1c _3-s_1 s_2 s_3 & -s_1 c_2 & -s_1 s_2 c_3-s_3c _1 \\
    		s_1c _3+s_2 s_3c _1 &c _1c _2 & s_2c _1c _3-s_1 s_3 \\
    		s_3c _2 & -s_2 &c _2c _3 \\
    	\end{array}
    	\right)
    \end{equation}
where $c_i = \cos{{\alpha_i}}$ and $s_i =\sin{{\alpha_i}}$, $i=1,2,3$. Note that this form does not include 'reflections', i.e. those $\Omega$ for which $\det(\Omega)=-1$\footnote{These transformations create, up to the sign of its individual components, the same forms of the matrix $M_D$ as rotations.}. Further we will analyze the case when the $\Omega$ matrix is real. This, in turn, determines the type of the Dirac mass matrix
    \begin{eqnarray}
        M_{D} = i \, U_\nu \sqrt{\hat{m}} \Omega\sqrt{\hat{M}} U_N^\dagger
        \label{eq:md}
    \end{eqnarray}
Next, we will work in the mass basis of heavy neutral leptons $N_1,N_2,N_3$, i.e. $U_N=I$. The observables, namely, the lifetime and the fraction of energy considered in Section \ref{sec:cosmology}, contain $\Theta_{\alpha I}$ (index $I=1,2,3$ numbers the mass states of HNL, $\alpha= e, \mu, \tau$) and obviously do not depend on the choice of $U_N$
    \begin{equation}
        \label{rv:general}
        \Theta_{\alpha I} = \frac{i\sum_{k}(U_\nu)_{\alpha k}\sqrt{m_k}~\Omega_{k I}^*}{\sqrt{M_I}}
    \end{equation}
Assuming that all angles of $\Omega$ matrix are real we get a constraint for the form \eqref{rv:general} with real $\Omega$. Take all the components of $U$ and $\Omega$-matrices bounded from above, namely $|U_{\alpha k}|\leq 1$, $|\Omega_{k I}| \leq 1$, and also take into account that $\sqrt{m_i} < \sqrt{\sum_{k=1}^3m_k}$
    \begin{eqnarray}
        \label{eq:bounds:general}
        \sum_\alpha |\Theta_{\alpha1}|^2\Big|_{\Omega \text{-real matrix}} < \sum_\alpha \left(\frac{\sum_{k}\sqrt{\Sigma_i m_i}}{\sqrt{M_I}}\right)^2 < 27\cdot10^{-3}\left(\frac{\overline{\Sigma m}}{1~\mbox{eV}}\right)\left(\frac{M_{1}}{1\mbox{~keV}}\right)^{-1},
    \end{eqnarray}
where $\overline{\Sigma m}$ is the upper bound limit for the sum of the masses of active neutrinos.
Consider the case of a normal hierarchy for active neutrinos ($m_3\gg m_1\sim m_2$) and a hierarchy for HNL corresponding to the model $\nu$MSM ($M_1\ll M_2\approx M_3 =M$), then $\hat{M}\approx diag(0,M,M)$ and $\hat{m} \approx diag(0,0,m_3)$, and the matrix $M_D$ has the form
    \begin{equation}
        \label{md:2}
        M_D \approx i\sqrt{M} \sqrt{m_3}
        \left(
        \begin{array}{ccc}
        \mathcal{O}(\sqrt{\frac{M_1}{M}}) & \Omega_{32} U_{e3}^* & \Omega_{33} U_{e3}^* \\
        \mathcal{O}(\sqrt{\frac{M_1}{M}}) & \Omega_{32} U_{\mu 3}^* & \Omega_{33} U_{\mu 3}^* \\
        \mathcal{O}(\sqrt{\frac{M_1}{M}}) & \Omega_{32} U_{\tau 3}^* & \Omega_{33} U_{\tau 3}^* \\
        \end{array}
        \right),
    \end{equation}
so the spectrum of eigenvalues leads to the implementation of the Yukawa hierarchy 
    \begin{equation*}
        \hat{F} = \frac{1}{v}~diag\left(\sim 0,~\sim 0,~ i\sqrt{M}\sqrt{m_3}(\Omega_{32} U_{\mu3}+\Omega_{33} U_{\tau3})\right),
    \end{equation*}
that for $\Omega_{33}=c_2c_3=1$ is consistent with the simplest and convenient case $\Omega=I$ ("minimal parametric mixing"). However, if $\Omega_{33}=\Omega_{32}=0$,~i.e. $\alpha_2=0$ and $\alpha_3=\frac{\pi}{2}$, which correspond to the case when all Yukawa constants have the same order of smallness. Consider this special case of choosing $\alpha_1$ and $\alpha_2$ parameters. As we can see from \eqref{eq:omega:asym} the choice of $\Omega=I$ is not suitable 
    \begin{equation}
        \label{eq:omega:asym}
        \Omega(\alpha_1,0,\pi/2) = \left(
            \begin{array}{ccc}
             0 & -s_1 & -c_1 \\
             0 & c_1 & -s_1 \\
             1 & 0 & 0 \\
            \end{array}
        \right),
    \end{equation} 
but reduction to the case $\Omega = I$ can be obtained by changing the basis of HNLs $N_I^\prime = V N_I$. In fact, the identical transformation of the expression \eqref{eq:md} gives
    \begin{equation}
        M_D = i U_\nu\sqrt{\hat{m}}\Omega \underbrace{U_N^{\dagger} U_N}_{=I} \sqrt{\hat{M}}U_N^{\dagger} = iU_\nu\sqrt{\hat{m}}\widetilde{\Omega} \sqrt{\widetilde{M}},
    \end{equation}
where $\sqrt{\widetilde{M}} = U_N\sqrt{\hat{M}}U_N^\dagger$ and $\widetilde{\Omega}=\Omega U_N^\dagger$, so the same matrix $M_D$ corresponds to a different mass ordering for HNL and $\Omega$ matrices.
If one takes the rotation matrix as
    \begin{equation}
        U_{rot} = \left(
        \begin{array}{ccc}
        0 & 0 & -1 \\
        0 & 1 & 0 \\
        1 & 0 & 0 \\
        \end{array}
        \right),
    \end{equation}
    \begin{equation}
        \widetilde{\Omega}_{(asym)} = \Omega_{(asym)} U_{rot}^\dagger = \left(
        \begin{array}{ccc}
        c_1 & -s_1 & 0 \\
        s_1 & c_1 & 0 \\
        0 & 0 & 1 \\
        \end{array}
        \right)
    \end{equation}
then, taking $\alpha_1=0$, we get the case $\widetilde{\Omega}=I$ with a modified ordering of HNL masses $\widetilde{M}=diag(M_3,M_2,M_1)$ instead of the original one $\hat{M}=diag(M_1,M_2,M_3)$.
Thus, the same Lagrangian \eqref{lagr} corresponds to different HNL mass orderings, and in some special cases to reproduce the convenient case of "minimal parametric mixing" $\Omega=I$ we need a mass reordering in the heavy leptons sector or, equivalently, a transformation of the basis for HNL mass states using a specific $U_N$.

By choosing $\Omega = I$, we get a view of the mixing matrix between sterile and active neutrinos that does not contain additional unknown complex angles
\begin{eqnarray}\label{rv:explicit}
    \Theta = \left( \begin{array}{ccc}
    iU_{e1}\sqrt{\frac{m_{1}}{M_{1}}} & iU_{e2}\sqrt{\frac{m_{2}}{M_{2}}} & iU_{e3}\sqrt{\frac{m_{3}}{M_{3}}} \\
    iU_{\mu1}\sqrt{\frac{m_{1}}{M_{1}}} & iU_{\mu2}\sqrt{\frac{m_{2}}{M_{2}}} & iU_{\mu3}\sqrt{\frac{m_{3}}{M_{3}}} \\
    iU_{\tau1}\sqrt{\frac{m_{1}}{M_{1}}} & iU_{\tau2}\sqrt{\frac{m_{2}}{M_{2}}} & iU_{\tau3}\sqrt{\frac{m_{3}}{M_{3}}}
    \end{array} \right) 
\end{eqnarray}
In the following this explicit form of the mixing matrix is denoted as $\overline{\Theta}$.

A discussion of the ambiguity of choosing the $\Omega$ matrix in the case of supersymmetric models can be found in \cite{ibarra2}. In the SM with a massless neutrino, the Yukawa matrix in the charged leptons sector and the matrix of gauge interactions of bosons with fermions are diagonal, so the lepton flavor is conserved. Non-zero neutrino masses and neutrino mixing lead to a lepton flavor violation (LFV) by analogy with CKM mixing for quarks. The completely unobservable level of LFV in the SM due to the small neutrino masses \cite{small_lfv} can be enhanced by HNL lagrangian terms and soft supersymmetry breaking terms, providing new contributions to LFV processes, such as $\mu\to e\gamma$. The supersymmetric see-saw mechanism is stable with respect to radiative corrections due to the presence of right sneutrinos, which suppress large HNL corrections to the Higgs boson mass, however, a large degree of freedom remains for determining Yukawa matrices in the neutrino and charged lepton sectors, which would allow one to calculate the LFV decay channels unambiguously. There is no reason to assume that the Yukawa matrices for charged leptons and $M_D$ for the neutrino sector are simultaneously diagonal. In the basis where the charged-lepton Yukawa matrix and the gauge boson interactions are flavor-diagonal, one can use 
\eqref{eq:md}, depending on the experimentally measured parameters for active neutrinos, three masses of heavy neutral leptons and three parameters defining $\Omega$.

The "minimal parametric mixing" $\Omega=I$ implies that one can work in a basis where $M_D$ and $M_N$ are both diagonal \cite{hisano}. One can observe by looking at the currents (\ref{lagr:act currents:cc})-(\ref{lagr:act currents:nc}) and
\eqref{eq:md} that such a basis can be constructed by rotating the left charged leptons with the matrix $U_\nu$, then LFV occurs in the charged sector. An alternative choice of the $\Omega$ matrix, a special case of which is described above, uses the form $M_D=X\hat{N}$, where $X$ is a unitary matrix, and $\hat{N}$ is a diagonal matrix \cite{ibarra2}. In this case, it follows from 
\eqref{eq:seesaw} that $\hat{M} = -\hat{N}X^Tm^{-1}_\nu X\hat{N}$, which allows simultaneously diagonalizing sectors of active neutrinos and charged leptons when the basis $N_i$ of the right (sterile) neutrinos is rotated by the $X$-matrix. The mass matrix $M_N$ of sterile neutrinos in this case is non-diagonal. Such a choice of $\Omega$ can describe scenarios in which LFV occurs in the HNL sector. These specific cases do not exhaust all possible options for choosing $\Omega$.

In order to control the numerical stability of the $\mathcal{O}(R^{2})$ diagonalization procedure, it is convenient to encode the model in the LanHEP \cite{lanhep} format. For a given physical masses, the LanHEP package calculates the mass parameters in the Lagrangian analytically, and the mass matrix is then diagonalized by orthogonal transformation using numerical methods \footnote{using the procedure contained in CERNlib 202}. For example, for HNL masses $M_{1} = 10$ keV, $M_{2}=400$ MeV, $M_{3}=420$ MeV with a large splitting corresponding to the exclusion contours for experiments on colliders \cite{gorbunov}, the non-diagonal elements of the mass matrix are less than 10$^{-6}$ MeV, at the same time, the given masses and the numerically calculated ones coincide with high accuracy.


\section{Cosmological restrictions for heavy neutral leptons as warm dark matter}
\label{sec:cosmology}
In the following we assume that the heavy neutral leptons $N_{1,2,3}$ are ordered by mass, and $N_1$ is the lightest of them. On the scale of $M_1\sim$keV, the main decay channel of HNL is the decay into three neutrinos $N_1\to \nu\nu\nu$ (Dirac basis), the decay width corresponding to the effective four-fermion amplitude determined by 
\eqref{lagr:hnl currents} can be written as
\begin{equation}
    \label{eq:width1}
    \Gamma\Big(N_1\rightarrow \sum_{\alpha,\beta}\nu_\alpha,\nu_\beta,\overline{\nu_\beta}\Big) = \frac{G_F^2M_{1}^5}{192\pi^3}\sum_\alpha |\Theta_{\alpha1}|^2,
\end{equation}
where $\alpha,\beta = e,\mu,\tau$. Given the affiliation of neutrino to Majorana fermions, it is consistent to take into account the contribution from the conjugated decay channels\footnote{in the case of identical conjugate state for Majorana neutrinos factor 2 appears in the amplitude calculation}
\begin{eqnarray}
    \label{eq:width2}
    \Gamma_{N_1 \rightarrow \nu\nu\nu} &=& \Gamma\Big(N_1\rightarrow \sum_{\alpha,\beta}\nu_\alpha,\nu_\beta,\overline{\nu_\beta}\Big)
    + \Gamma\Big(N_1\rightarrow \sum_{\alpha,\beta}\overline{\nu_\alpha},\overline{\nu_\beta},\nu_\beta\Big).
\end{eqnarray}

\begin{figure}[t!]
        \centering
        \includegraphics[scale=0.6]{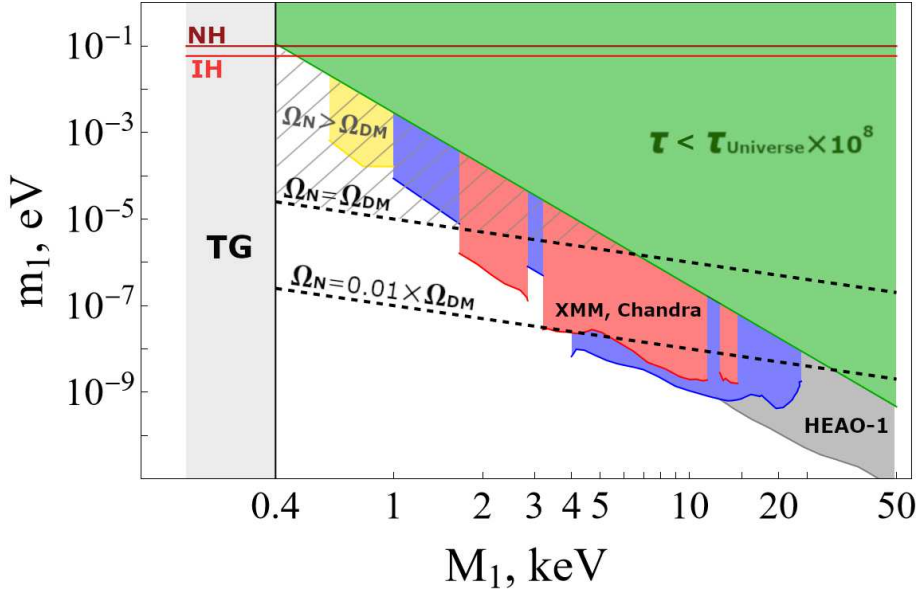}
        \caption{\small Restrictions on the mass of the lightest active neutrino depending on the mass of the HNL particle of dark matter in the case of "minimal parametric mixing" $\Omega=$I. The green area shows the universal limit from the gamma astronomical observations $\tau_N > 10^{25}$ seconds. The light gray area is the Tremaine-Gunn bound (TG). The red and dark red horizontal lines correspond to the data on neutrino oscillations in the case of a normal (NH) and inverse (IH) hierarchy of active neutrino masses. The red, blue, gray and yellow regions are excluded by various gamma astronomical observations and correspond to \cite{gamma_astro}. The dashed lines indicate the area in which the density of dark matter is realized by the Dodelson-Widrow mechanism \cite{DW}, the upper bound corresponds to 100$\%$ of the $N_1$ dark matter density, the lower one is 1$\%$.}
        \label{fig:1}
    \end{figure}

Further, we take into account the 0.4 keV estimate from below for the HNL mass, since the distribution of HNL as fermionic dark matter in the phase space of the galaxy is limited by the distribution for a degenerate fermi gas (Tremaine-Gunn bound, \cite{Tremaine:1979we}).
It is obvious that a sterile neutrino - a candidate for the role of dark matter - does not decay on cosmological time scales of the order of the age of the Universe, which means $\tau_{N_1}\ge 4\times10^{17}\mbox{~seconds}$. The secondary one-loop radiative decay $N\rightarrow\gamma,\nu$, which can be a distinctive signal with photon energy $E_\gamma=M_1/2$, has the width
    \begin{equation}
        \Gamma\Big(N_1 \rightarrow \gamma,\nu\Big) = \frac{9 \alpha_{EM} G_F^2 M_1^5}{256 \pi^4} \sum_\alpha |\Theta_{\alpha1}|^2.
    \end{equation}
Although $\Gamma_{N\rightarrow \nu\nu\nu}/\Gamma_{N\rightarrow \gamma\nu}\equiv k_{rad}= \frac{8\pi}{27\alpha_{EM}} \approx 128$ and radiation decay does not make significant corrections to the lifetime by changing it by a factor of $k_ {rad}/(1+k_{rad})\sim 1$, the lifetime limit can be increased by 8 orders of magnitude due to the specifics of gamma-ray astronomical observations, see \cite{gamma_astro,xray1,xray2}. In the following the estimate $\tau_{N_1} > 10^{25}$ seconds will be used.
Lifetime of the lightest HNL in seconds
    \begin{equation}
        \label{eq:lifetime_sec}
        \tau_{N_1}= 2.88 \times 10^{19}\left(\frac{M_1}{1\mbox{~keV}}\right)^{-5}\Big(\sum\limits_{\alpha}{|\Theta_{\alpha1}|^2}\Big)^{-1} \mbox{~sec}.
    \end{equation}
Note that the elements of the $\Omega$ matrix into $\Theta_{\alpha 1}$ are represented by "weights" of contributions of the each active neutrino masses. 
The case of "minimal parametric mixing" gives a restriction directly on the mass $m_1$.
    \begin{equation}
        \label{eq:bounds:explicit}
        \sum_\alpha |\overline{\Theta}_{\alpha 1}|^2 = \sum_\alpha \frac{m_1}{M_{1}} |U_{\alpha 1}|^2 = \frac{m_1}{M_1}.
    \end{equation}
Here we used the unitarity of the PMNS matrix, assuming that $U_\nu \simeq U_{\rm PMNS}$ up to $\mathcal{O}(R^2)$.
Then from \eqref{eq:lifetime_sec} taking into account \eqref{eq:bounds:explicit} one can get
    \begin{equation}
        \label{eq:summ:bound}
        \left(\frac{m_1}{1\mbox{~eV}}\right) < \frac{2.88\times 10^{22}}{\tau_{X}(\mbox{sec})}\left(\frac{M_1}{\mbox{1~keV}}\right)^{-4},
    \end{equation}
where $\tau_{X}$ are the current limits on the lifetime of dark matter from gamma - astronomical observations \cite{gamma_astro}.
If the active neutrinos are ordered according to the inverse mass hierarchy (IH) then $m_1 \simeq \Sigma_i m_i$. But when there is a normal mass hierarchy (NH) for active neutrinos the small scale of $m_1$ is not associated with the data of neutrino oscillations: $\Sigma_i m_i > \sqrt{\Delta m^2_{atm}+\Delta m^2_{sol}}+\sqrt{\Delta m^2_{sol}} \approx 0.099$ eV (NH), $\Sigma_i m_i > \sqrt{\Delta m^2_{atm}+\Delta m^2_{sol}}+\sqrt{\Delta m^2_{sol}}\approx 0.058$ eV (IH). In case of IH there is a negligibly small allowed parameter area near the TG bound and above the red horizontal line for IH (see Fig.\ref{fig:1}). Hence, it is almost excluded by the "minimal parametric mixing" scenario. 
To estimate the permissible deviation of the matrix $\Omega$ from the unit matrix one can rewrite the mixing parameter in the form
    \begin{multline}
        \label{eq:omega:non-diagonal}
        \sum\limits_\alpha |\Theta_{\alpha1}|^2 = \frac{1}{M_1} \sum\limits_\alpha \Big|\sum\limits_j\Omega_{j1} U_{\alpha j} \sqrt{m_j}\Big|^2 \simeq
        \\
        \simeq\frac{3 m_3}{M_1}\left(\Omega_{11}\sqrt{\frac{m_1}{m_3}} + \Omega_{21}\sqrt{\frac{m_2}{m_3}} + \Omega_{31}\right)^2,
    \end{multline}
where all elements of the matrix $U_{\rm PMNS}$ are set equal to 1, which is a reasonable estimate in order of magnitude.
Then from the lifetime constraints of $\tau_N > 10^{25}$ seconds and \eqref{eq:omega:non-diagonal} we get
\begin{equation}
    \label{xxx}
    \left(\Omega_{11}\sqrt{\frac{m_1}{m_3}} + \Omega_{21}\sqrt{\frac{m_2}{m_3}} + \Omega_{31}\right)^2
    \le 10^{-3}\left(\frac{m_3}{1\mbox{~eV}}\right)\left(\frac{M_1}{1\mbox{~keV}}\right)^{-4}.
\end{equation}

\begin{figure}[!t]
        \centering
        \includegraphics[scale=0.5]{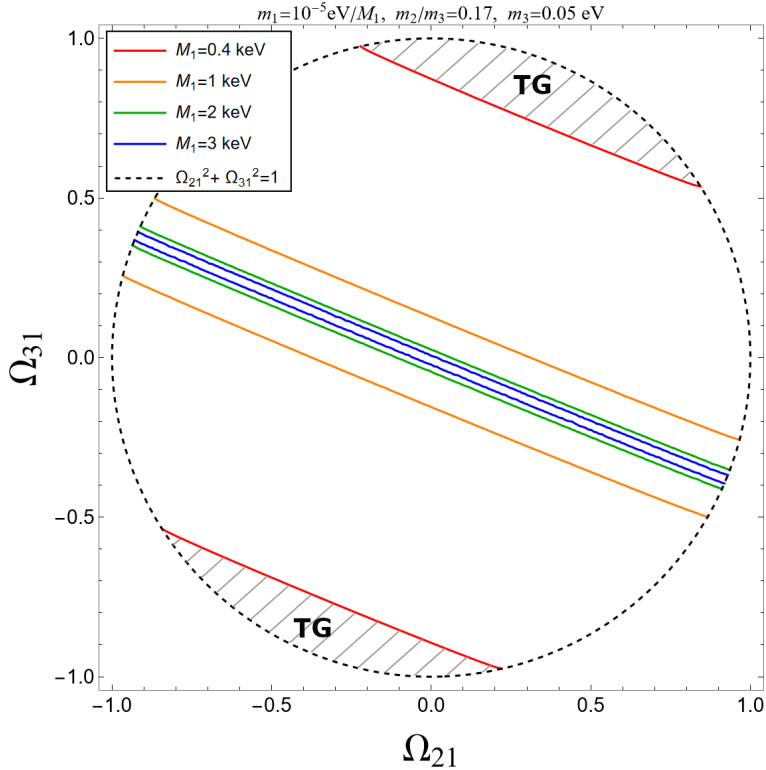}
        \caption{The ranges of values of the non-diagonal elements $\Omega_{21}$ and $\Omega_{31}$ of the first column of the real matrix $\Omega$ satisfying the lifetime limit of the light HNL of dark matter \cite{gamma_astro} in the case of a normal hierarchy of active neutrino masses. Estimates of \eqref{eq:energy3} for the parameter $m_1$ are taken into account in the case when a sterile neutrino realizes $100\%$ of the density of dark matter ($m_1=10^{-5}/(M_1[\text{keV}]$) eV). The dotted circle corresponds to the limiting case $\Omega_{11} = 0$. Solid lines correspond to the boundaries of the allowed area, the allowed area itself lies between them and inside the circle. Shaded areas are excluded by the Tremaine-Gunn boundary (TG).}
        \label{fig:2}
    \end{figure}

To plot the contours, we use the following estimate for the masses of active neutrinos $m_1= 10^{-5} \left(\frac{M_1}{1\mbox{~keV}}\right)^{-1}$eV, $m_2=\sqrt{\Delta m^2_{sol}}$ and $m_3 \simeq \sqrt{\Delta m^2_{atm}}$. In addition, we assume that the elements of the matrix $\Omega$ are real. The corresponding contour for direct mass hierarchy is shown in Fig.\ref{fig:2}. One can see that when the HNL DM mass increases, the band of allowed values is narrowing, and the domain for the parameter $\Omega_{31}$ tightens more significantly, since in the direct hierarchy the $\upnu_3$ state is the heaviest one. Localization of the permissible parameters near the center of the circle (i.e. to the "minimal mixing") does not occur completely due to the resulting reductions in the contributions of terms with $m_2$ and $m_3$.
    
The cosmological restriction in the HNL sector for the fraction of energy in the Universe appears within the framework of the picture where $\Theta$ is small enough, and the sterile neutrino has never been in thermal equilibrium. The dominant mechanism of the formation of sterile neutrinos (Dodelson-Widrow mechanism \cite{DW}, also \cite{Dolgov:2000ew}) is due to oscillations of active and sterile neutrinos. The proportion of energy in the Universe in the case of non-resonant production \cite{Abazajian:2001nj,Viel} is given by the formula
    \begin{equation}
        \label{eq:energy1}
        \Omega_Nh^2 \sim 0.1 \sum\limits_{I=1}^{3}\sum\limits_{\alpha=e,\nu,\tau}\left(\frac{|\Theta_{\alpha I}|^2}{10^{-8}}\right)\left(\frac{M_I}{1\mbox{~keV}}\right)^2.
    \end{equation}
Using $\Theta=M_D M_N^{-1}$ with $U_N=I$, one can write
    \begin{eqnarray}
        \label{eq:energy2}
        \sum\limits_{I,\alpha} |\Theta_{\alpha I}|^2 = Tr\left(M_D\hat{M}^{-1} (\hat{M}^{-1})^\dagger M_D^\dagger\right)
        = \sum\limits_{I, \alpha} |(M_D)_{I\alpha}|^2 \cdot\frac{1}{M_I^2}.
    \end{eqnarray}
Then it follows from \eqref{eq:energy1} together with \eqref{eq:energy2} that if $N_1,...,N_\mathcal{K}$ are HNL dark matter particles, then
    \begin{equation}
        \label{eq:energy3}
        \sum\limits_{I=1}^{\mathcal{K}\leq3}\sum\limits_\alpha |(M_D)_{\alpha I}|^2=m_0^2~,~m_0=\mathcal{O}(0.1\mbox{~eV~}).
    \end{equation}
Note that the value of $m_0$ does not depend on the values of the masses of $M_I$. In the case of "minimal parametric mixing" $(M_D)_{\alpha I}=iU_{\alpha I}\sqrt{m_I}\sqrt{M_I}$ and the condition \eqref{eq:energy3} immediately gives the mass scale
    \begin{equation}
        \label{eq:energy:m1}
        m_1 = \frac{m_0^2}{M_1}= 10^{-5}\mbox{~eV}\times\left(\frac{M_1}{1\mbox{~keV}}\right)^{-1}.
    \end{equation}

Note that this estimate can also be obtained in the general case without fixing $\Omega=I$, see \cite{shaposhnikov1}. More details related to \eqref{eq:energy:m1} are given in the Appendix.

\section{Summary}
\label{sec:conclusions}
The ambiguity of the choice of the mixing matrix in the extended lepton sector can play a role for the observed consequences. The most significant practical result of a consideration above that goes beyond the simplified 'model-independent phenomenological approach', is the matrix of "minimal parametric mixing", Eq.\ref{rv:explicit}, diagonalizing the mass term of the lepton sector for three generations and containing in the sector of neutral heavy leptons (Majorana neutrinos) only dependence on their masses with an apparent hierarchy of values of matrix elements. Such mixing corresponds to the simplest choice of $\Omega=I$ for the matrix of "weights" (Section 3) or its equivalents corresponding to the reordering of the masses of neutral heavy leptons. "Minimal parametric mixing" immediately leads to a restriction from above $m^2_0/M_1$ for the lightest active neutrino (the known parameter of dark matter $m^{dm}_\nu=m^2_0/M_1$ is of the order of 10$^{-5}$ eV on the mass scale of $M_1$ of the order of keV, see \cite{shaposhnikov1}). Other representations for the matrix of "weights" $\Omega$ correspond to more complex functional forms of the mixing matrix, when the first column contains not only the contributions of the lightest active neutrino. A fairly accurate estimation for the non-diagonal elements of the $\Omega$ matrix by order of magnitude based on the lifetime constraints from below for the lightest sterile lepton leads to the need for their suppression by two orders of magnitude or more compared to one, in a strong dependence of suppression on the mass of the lightest HNL - a candidate for the role of dark matter.

In addition it is interesting to note that neutrino mass generation by means of the see-saw mechanism has been analysed in connection with the minimal supersymmetry, since in this case the see-saw mechanism is stable with respect to radiative corrections when pronounced neutral heavy lepton mass hierarchy is assumed. Most notable experimental indications in such a framework are lepton flavor violating processes, in particular $\mu \to e \gamma$ \cite{ibarra2}, \cite{hisano}, \cite{emugamma,emugamma1,emugamma2,emugamma3,emugamma4}. It follows from these considerations that nonobservation or potential observation of LFV allows one to identify three preferential scenarios in the extensions of leptonic sector by HNL, namely 
(i) there exists a basis where the active neutrino Yukawa matrix $M_D$ and the HNL matrix $M_N$ are simultaneously diagonal, the charged lepton Yukawa matrix is not diagonal and LFV takes place in the sector of charged standard leptons. In this "minimal parametric mixing" with $\Omega = 1$ Br($\mu \to e \gamma$) may be considerably higher than the SM value exceeding the available experimental limits; 
(ii) there exists a flavor basis where the charged lepton Yukawa matrix and the active neutrino Yukawa matrix are simultaneously diagonal, while the HNL mixing martix is not diagonal, so LFV is attributed to the sterile neutrino sector. Then Br($\mu \to e \gamma$) is negligibly small. In this scenario the active neutrino - HNL mixing matrix $\Theta$ can be reduced to the minimal parametric form by reordering of the HNL masses; 
(iii) a scenario when the active neutrinos do not have a clearly defined hierarchy of masses, then Br($\mu \to e \gamma$) is suppressed by a mechanism analogous to Glashow–Iliopoulos–Maiani (GIM) for a real $\Omega$ matrix. In the case of a neutrino mass spectrum with a small splitting at the keV scale which we are mainly considering in the $\nu$MSM model, the supersymmetric renormalization group corrections that develop off-diagonal contributions to the leptonic soft supersymmetry breaking terms do not play a critical role in the absence of separated HNL mass scales. However, if available, they can influence significantly on the LFV restrictions.

Acknowledgments: M.D. is grateful to Mikhail Shaposhnikov for useful discussion. The research was carried out within the framework of the scientific program of the National Center for Physics and Mathematics, project "Particle Physics and Cosmology".\\

{\large \bf Appendix 1}\\

In this Appendix we reproduce the general restriction for the number of dark matter HNLs \cite{shaposhnikov1} in detail.
One can write the diagonal see-saw matrix as the sum 
\begin{equation}
\hat{m} = S_1+S_2+S_3 
\label{sum}
\end{equation} 
with $S_I$ parametrized as
\begin{equation}
    (S_I)_{ij}= X_{Ii}X_{Ij} = 
\left( \begin{array}{ccc}
 X_{I1}^2 & X_{I1} X_{I2} & X_{I1} X_{I3} \\
 X_{I2} X_{I1} & X_{I2}^2 & X_{I2} X_{I3} \\
 X_{I3} X_{I1} & X_{I3} X_{I2} & X_{I3}^2 
\end{array} \right),
\end{equation}
where
\begin{equation}
    X_{Ii} = \frac{(M_DU_\nu)_{Ii}}{\sqrt{M_I}}.
\end{equation}
One can check that $\det{S_I}=0$. In terms of $X$-matrix the condition \eqref{eq:energy1} looks as
\begin{equation}
    \sum\limits_{I,\alpha} \frac{M_{I}}{M_1}|X_{I \alpha}|^2 = \frac{m_0^2}{M_1} \equiv m_\nu^{dm},
\end{equation}
so if $M_1$ is of the order of keV the dark matter parameter $m_\nu^{dm}=\mathcal{O}(10^{-5}$eV). This bound
is independent on the explicit form of the mixing. In the case of $\mathcal{N}=3$ heavy neutral leptons, taking trace of \eqref{sum} leads to
\begin{eqnarray*}
    m_1+m_2+m_3 = \sum\limits_{I=1}^{\mathcal{N}}\sum\limits_{i=1}^{3} Re[X_{Ii} X_{Ii}] \leq \sum\limits_{I=1}^{\mathcal{N}}\sum\limits_{i=1}^{3} |X_{Ii}|^2 \leq \\ \leq \sum\limits_{I=1}^{\mathcal{N}}\sum\limits_{i=1}^{3} \frac{M_I}{M_1}|X_{I \alpha}|^2 = m_\nu^{dm},
\end{eqnarray*}
where $M_1<M_2,M_3$ and obviously for any complex $x$ $Re[x^2] \leq |x|^2$. It follows that for $\mathcal{N}=3$
HNL's
\begin{equation}
    m_1+m_2+m_3 \leq m_\nu^{dm} = \mathcal{O}(10^{-5}{~eV}),
\end{equation}
which is excluded. In the case when only $N_1$ and $N_2$ are the dark matter HNL's, 
\begin{equation}
    m_1 + m_2 + m_3 \leq m_\nu^{dm} + \sum\limits_{i=1}^{3}{Re}[X_{3i}].
\end{equation}
According to this expression it is easy to show that $\sum\limits_{i=1}^{3}{Re}[X_{3i}] > m_3$. Using $\det{(S_1 + S_2)}=\det{(\hat{m} - S_3)}=0$ we can write in component notation
\begin{eqnarray*}
 \det \left( \begin{array}{ccc}
 m_{1}-X_{31}^2 & -X_{31} X_{32} & -X_{31} X_{33} \\
 -X_{31} X_{32} & m_{2}-X_{32}^2 & -X_{32} X_{33} \\
 -X_{31} X_{33} & -X_{32} X_{33} & m_{3}-X_{33}^2 
\end{array} \right) =\\= m_{1} m_{2} m_{3}-m_{1} m_{2} X_{33}^2-m_{1} m_{3} X_{32}^2-m_{2} m_{3} X_{31}^2 = 0
\end{eqnarray*}
or, taking the real part, we come to a contradiction
\begin{eqnarray*}
    1 = \frac{Re[X_{33}^2]}{m_3} + \frac{Re[X_{32}^2]}{m_2} + \frac{Re[X_{31}^2]}{m_1} > \frac{\sum X_{3i}^2}{m_3} >  1 ,
\end{eqnarray*}
so the case when two HNLs are the dark matter particles does not respect \eqref{eq:energy1}. Only the case of a single $N_1$ as the dark matter particle remains. It follows from  $\det{(\hat{m} - S_1})=0$ that
\begin{eqnarray*}
    m_1=X_{11}^2 + \frac{m_1}{m_2} X_{12}^2 + \frac{m_1}{m_3} X_{13}^2 < \sum\limits_{i} X_{1i}^2,
\end{eqnarray*}
\begin{eqnarray*}
    m_1 < \sum\limits_{i} Re(X_{1i}^2) < \sum\limits_{i} |X_{1i}|^2 < \sum\limits_{i} \frac{M_I}{M_1} |X_{1i}|^2 = m_\nu^{dm}
\end{eqnarray*}
so finally $m_1 < m_\nu^{dm}=\mathcal{O}(10^{-5}{~eV~})$.


\end{document}